\documentclass{pazh}
\usepackage{float,epsfig,psfig}
\usepackage{pstricks}

\begin{document}

\journalinfo{2002}{28}{4}{223}[236] % - fake info
\submitted{Received October 2, 2001}
\title{Searches for the Shell Swept up by the Stellar Wind from Cyg~OB2}

\author{T. A. Lozinskaya$^{1}$, V.V.Pravdikova$^{1}$,  A.V. Finoguenov$^{2}$}
\address{{$^1$ Sternberg Astronomical Institute, Universitetskii pr. 13,
    Moscow, 119899 Russia}\\ 
{$^2$ Space Research Institute, Russian Academy of Sciences, ul. Profsoyuznaya 84/32, Moscow, 117810 Russia}}
% \affil{Sternberg Astronomical Institute, Universitetskii pr. 13,
% Moscow, 119899 Russia}
% 
% \author{V.V.Pravdikova }
% \affil{Sternberg Astronomical Institute, Universitetskii pr. 13,
% Moscow, 119899 Russia}
% 
% \author{A.V. Finoguenov}
% \affil{Space Research Institute, Russian Academy of Sciences, ul.
% Profsoyuznaya 84/32, Moscow, 117810 Russia}
%\email{lozinsk@sai.msu.ru}

\maketitle

\begin{pazhabstract}
{\bf Abstract --} We investigated the kinematics of ionized gas in an
extended ($20^\circ\times15^\circ$) region containing the X-ray Superbubble
in Cygnus with the aim of finding the shell swept up by a strong wind from
Cyg~OB2. H$\alpha$~observations were carried out with high angular and
spectral resolutions using a Fabry--Perot interferometer attached to the
125-cm telescope at the Crimean Observatory of the Sternberg Astronomical
Institute.  We detected high-velocity gas motions, which could result from
the expansion of the hypothetical shell at a velocity of
25--50~km~s$^{-1}$. Given the number of OB~stars increased by Kn\"odlseder
(2000) by an order of magnitude, Cyg~OB2 is shown to possess a wind that is
strong enough [$Lw\simeq (1-2)10^{39}$~erg~s$^{-1}$] to produce a shell
comparable in size to the X-ray Superbubble and to a giant system of optical
filaments. Based on our measurements and on X-ray and infrared observations,
we discuss possible observational manifestations of the shell swept up by
the wind.


\end{pazhabstract}

\section{INTRODUCTION}

The studies of Cyg~OB2 by Reddish \emph{et al.} (1966) showed it to be a
Galactic association that is unique in compactness and in the density of massive
early-type stars. The question of where the shell swept up by the wind from this
compact grouping
of young stars is has been raised long ago. As such a shell, different authors
considered
the following: a system of optical filaments, 15$^\circ$ in size (Ikhsanov 1960;
Dickel
\emph{et al.} 1969); the inner (relative to it) system of filaments,
8--10$^\circ$ in
size, elongated across the Galactic plane (Bochkarev and Sitnik 1985); the
diffuse
component of the Cygnus~X radio source (Wendker 1970); the Cyg~4 or Cyg~5
gas--dust
shells from the list by Brand and Zealey (1975); and the giant X-ray Superbubble
in
Cygnus (Cash \emph{et al.} 1980). This list of possibilities is not yet
complete.

Recently, the data on the Cyg~OB2 stellar population have been radically
changed. Having
added stars previously hidden due to strong absorption, Kn\"odlseder (2000)
increased the
number of members and the total mass of the object by an order of magnitude. As
a result, the author classified this star grouping as a young globular cluster
similar to
the numerous blue globular clusters in the Large Magellanic Cloud (LMC), but it
is the
only one identified in the Galaxy so far. Currently, Cyg~OB2 numbers $\sim120$
O-type stars, which forces us to significantly (by an order of magnitude; see
Section~3)
increase the flux of ionizing radiation and the mechanical luminosity of the
stellar
wind and makes the searches for the swept-up shell around Cyg~OB2 even more
urgent.

\begin{figure*}[t!]
\centering\includegraphics[width=16.cm]{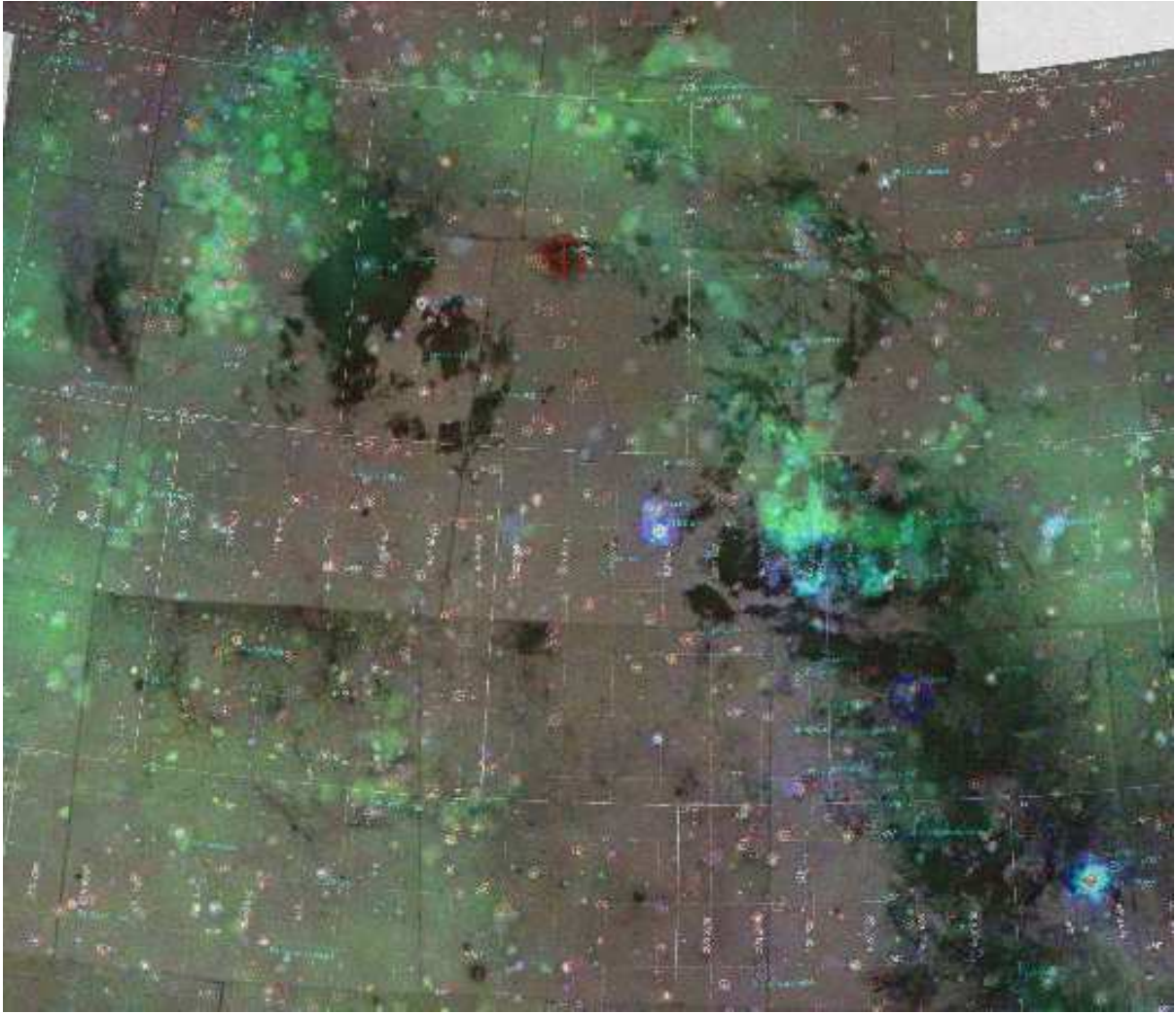}
\caption{%
Overall X-ray and optical structure of the Superbubble region in Cygnus. The
shades of blue-green indicate the X-ray (0.4--2.4~keV) brightness distribution
with a high
resolution obtained from the ROSAT archival data. The shades of dark gray
indicate an
optical photograph of the region (a mosaic of 19 Palomar Sky Survey red
prints).}%
%\label{%
%}%
\end{figure*}

The nature of the X-ray Superbubble in Cygnus and its relationship to Cyg~OB2
have been
discussed repeatedly. The observations published before~1984 were analyzed in
detail by
Bochkarev and Sitnik (1985). Uyaniker \emph{et al.} (2001) carried out new radio
continuum observations and comprehensively analyzed currently available
observational
data for the region in the X-ray and infrared bands and in molecular radio
lines. The
same conclusion was reached in the two studies: the Cygnus Superbubble is not a
physical
unity but is the sum of emissions from physically unrelated objects located at
distances
from~0.5 to 3.5~kpc.

Here, we pose the question from a different angle. There is no doubt that a
superposition of the X-ray, optical, infrared, and radio emissions from objects
located
at different distances along the spiral arm is observed in the Cygnus region.
Nevertheless, Cyg~OB2 is a compact star grouping that, probably, possesses the
strongest
stellar wind in the Galaxy whose action has continued for $\approx(2-3) 10^{6}$
yrs. In
that case, where is evidence for the action of this wind on the ambient
interstellar gas
in the first place? If none of the objects mentioned above can be identified as
the
shell swept up by the Cyg~OB2 wind, then what must be the parameters of the
ambient
interstellar medium for such a shell to be unobservable?

In attempting to detect the swept-up shell, we investigated the
radial-velocity field of the ionized gas in an extended
($20^\circ\times15^\circ$) region using our H$\alpha$ observations with a
Fabry--Perot interferometer attached to the 125-cm telescope at the Crimean
Observatory of the Sternberg Astronomical Institute (SAI).

In Section~1, the X-ray image constructed from the ROSAT archival data is
compared with an optical image of the entire region under study. Our
interferometric observations are presented in Section~2: we describe the
observing and data reduction techniques; construct the radial-velocity
distribution of the ionized gas in the region; and reveal the velocities
that could be attributable to an expansion of the shell swept up by the
wind. In Section~3, based on the possible expansion velocity of the
hypothetical shell we found and on currently available optical, X-ray,
radio, and infrared observational data, we attempted to find morphological
manifestations of this shell. Our conclusions are briefly summarized in the
last section.

\section{A COMPLETE X-RAY AND OPTICAL IMAGE OF THE REGION UNDER STUDY}

An image of the entire region under study, including the X-ray Superbubble in
Cygnus and
the giant system of optical nebulae, is shown in Fig.~1.

The high-resolution X-ray brightness distribution was constructed from the
ROSAT observational data. For imaging, we used archival data from the ROSAT
all-sky survey in three energy bands: 0.11--0.40, 0.40--1.0, and
1.0--2.4~keV. (The charged particle background was removed from these
images.)  Figure~1 shows the total emission in the three energy bands. Our
data reduction revealed the dominance of the 0.4--1.0~keV emission from the
Superbubble. The complete X-ray image for the $25^\circ\times 22^\circ$
region is a mosaic of 18 separate fields.

Optical emission distribution in the region superimposed on the x-ray is a
 mosaic map
constructed from 19 Palomar Sky Survey red prints. The optical emission is
represented
by the giant system of diffuse and thin-filament nebulae in the region with
$\mathrm{RA}=19^{\mathrm{h}} 30^{\mathrm{m}}\ldots
21^{\mathrm{h}}50^{\mathrm{m}}$,
 $\mathrm{D}=30\ldots47^\circ$; $l=70\ldots 88^\circ$, $b=-8\ldots +5^\circ$.

As Fig.~1 shows, the giant system of optical filaments and the X-ray Superbubble
as a whole closely coincide in the plane of the sky, considering that their
emission is produced by plasma with a distinctly different temperature.

\section{IONIZED-GAS VELOCITIES IN THE REGION}

\subsection*{Observations and Data Reduction}

In attempting to identify the shell produced by the Cyg~OB2 wind, we analyzed in
detail
the radial-velocity field by using the H$\alpha$ emission of the ionized gas in
this most
complex region of the Local spiral arm.

We measured the radial velocities in H$\alpha$ using a
Fabry--Perot interferometer with a focal reducer on the 125-cm
telescope at the Crimean Observatory of the SAI.

The observing program for this giant region was initiated in~1991. The detector
was an
image intensifier before~1994 and an ST-6 $242\times378$-pixel CCD array
after~1994.

An interference filter centered on H$\alpha$ with a FWHM of 25~\AA
was used to pre-monochromatize the emission. The actual spectral
resolution of the interferometer corresponded to
$\simeq15$~km~s$^{-1}$; the dispersion region (the
radial-velocity range free from an overlapping of adjacent
interference orders) was $\approx800$~km~s$^{-1}$. The [N~II]
6584~\AA line was in the middle of this range and was clearly
separated from H$\alpha$, whose largest zero-level width did not
exceed 200 km~s$^{-1}$ everywhere in the region under study.

The interferometer field of view and angular resolution for observations with
the focal
reducer at the Cassegrain focus of the 125-cm telescope are 10$'$ and 3--4$''$,
respectively.

A gas-filled tube was used as the laboratory H$\alpha$ source to calibrate the
radial
velocities and to allow for the instrumental profile of the interferometer.

The data reduction technique was detailed by Pravdikova (1995). Each line
profile was
fitted with one or more Gaussians by assuming that the FWHM of each component
was larger
than the FWHM of the instrumental profile and that the signal-to-noise ratio
was~$\ge 5$. The profile was fitted with a set of Gaussians in those cases where
either several
peaks, or a clear profile asymmetry, or broad line wings were observed.

 In the coarse of our  long-term program for studying the region, we
obtained more than 1000 interferograms. As a result of their
processing, we measured the velocities at maximum and the FWHMs
of individual H$\alpha$ components approximately at 20\,000
positions in the Cygnus region. All these data were used here.

Our observing technique allows us to investigate the extended complex in Cygnus,
more
than $20^\circ$ in size, with an angular resolution of 3--4$''$, but this
requires too
much observing time. Therefore, the region was covered with measurements with a
nonuniform density. Detailed observations with a multiple overlapping were
performed for
separate large areas. The velocity field was investigated most extensively in
the
central and western sectors of the region; in the rest of the region, mostly
brighter
nebulae (EM $\geq 100$~pc~cm$^{-6}$) have been investigated so far as a
tradeoff.  In
the areas of weak emission, we carried out mainly search observations with a
large step.
Only if anomalous velocities or multi-peaked line profiles were found did we
study the region in detail.

\begin{figure*}[t!]
\centering\includegraphics[width=14.cm]{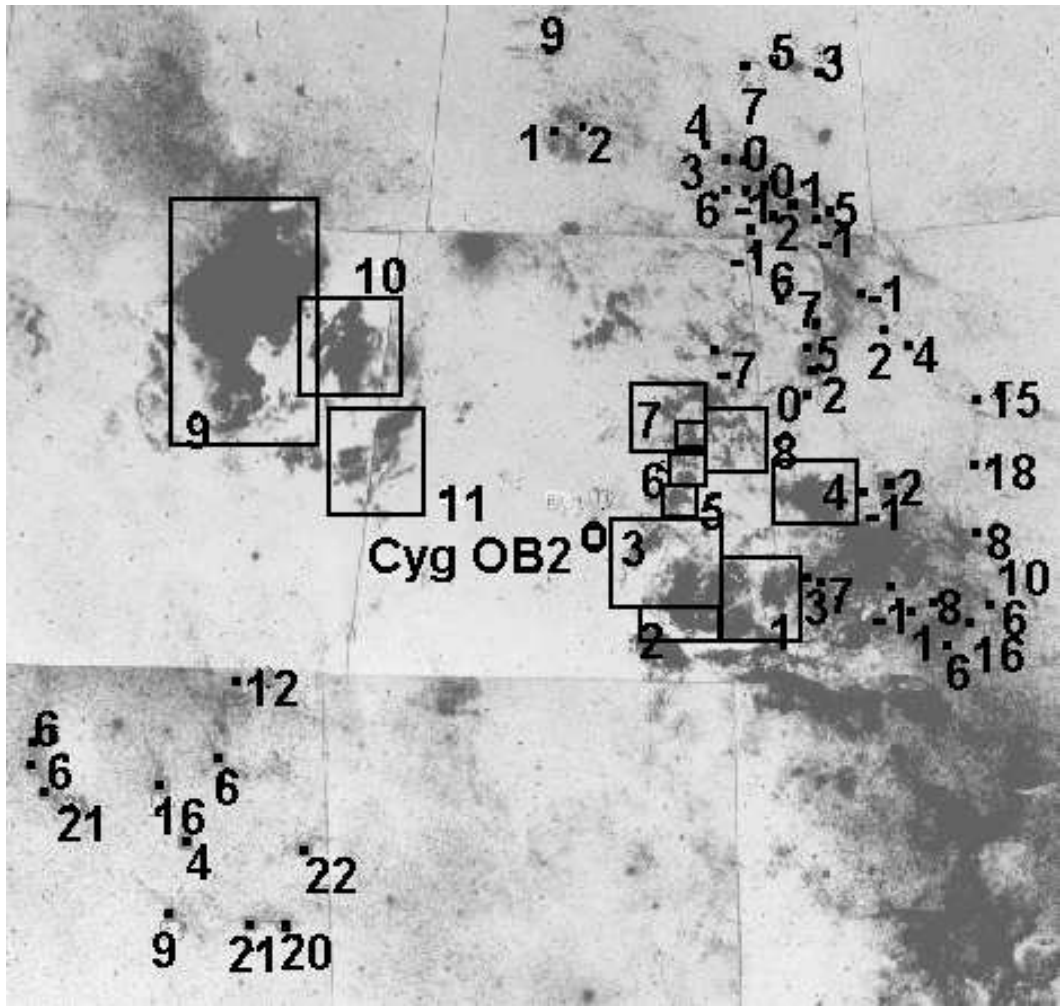}
%%% Figure:2
\caption{%
A map of optical emission in the region. The small heavy squares indicate the
localization of interference
H$\alpha$ rings for individual positions. The numbers beside the squares give
the radial velocities averaged over 10 to 20 measurements. The large rectangles
indicate the boundaries
of the fields for which the radial-velocity histograms were constructed from a
large number of H$\alpha$ measurements. (The corresponding histograms are shown
in Fig.~3).}%
%\label{%
%}%
\end{figure*}

The velocities and line widths for the extended filaments that form the giant
ring
system mentioned above were measured with poor spacial covering.

\subsection*{Results of the Interferometric Observations}

Figure~2 shows a map of optical emission and mark the regions with measured
radial
velocities.

The small heavy squares indicate the localization of the positions for which
individual
digital images were obtained with the Fabry--Perot etalon. The square size
roughly
corresponds to the size of an H$\alpha$ interference  ring in projection onto
the sky. The
numbers beside the squares give the radial velocities determined by processing
the
corresponding interferogram. Since the interference-ring radius is typically
$\approx
3-5'$, for our purposes, we disregarded the coordinate differences
between the
radial scans of the interference ring that were used to construct the line
profiles.
Accordingly, for each position, we averaged from 10--12 to 20--25 velocity
measurements
for different radial scans of one or two interference rings. The rms error of
each velocity measurement
varies from interferogram to interferogram within the range 1.5 to
4.5~km~s$^{-1}$. Only
for the group of very weak filaments in the southeast of the region does the
error
exceed these values, ranging from 5 to 9~km~$^{-1}$ for different
interferograms. In
this way, we present the results of our measurements for those positions at
which mostly
search observations were carried out.

The results of more detailed observations of the eleven fields for which many
interferograms that overlap the field with a large density were obtained are
presented in
the form of radial-velocity histograms. The rectangles in Fig.~2 mark the
boundaries of
fields nos.~1--11 for which radial-velocity histograms were constructed in
H$\alpha$. The
corresponding histograms are shown in Fig.~3.

\begin{figure*}[t!]
\centering\includegraphics[width=12.cm]{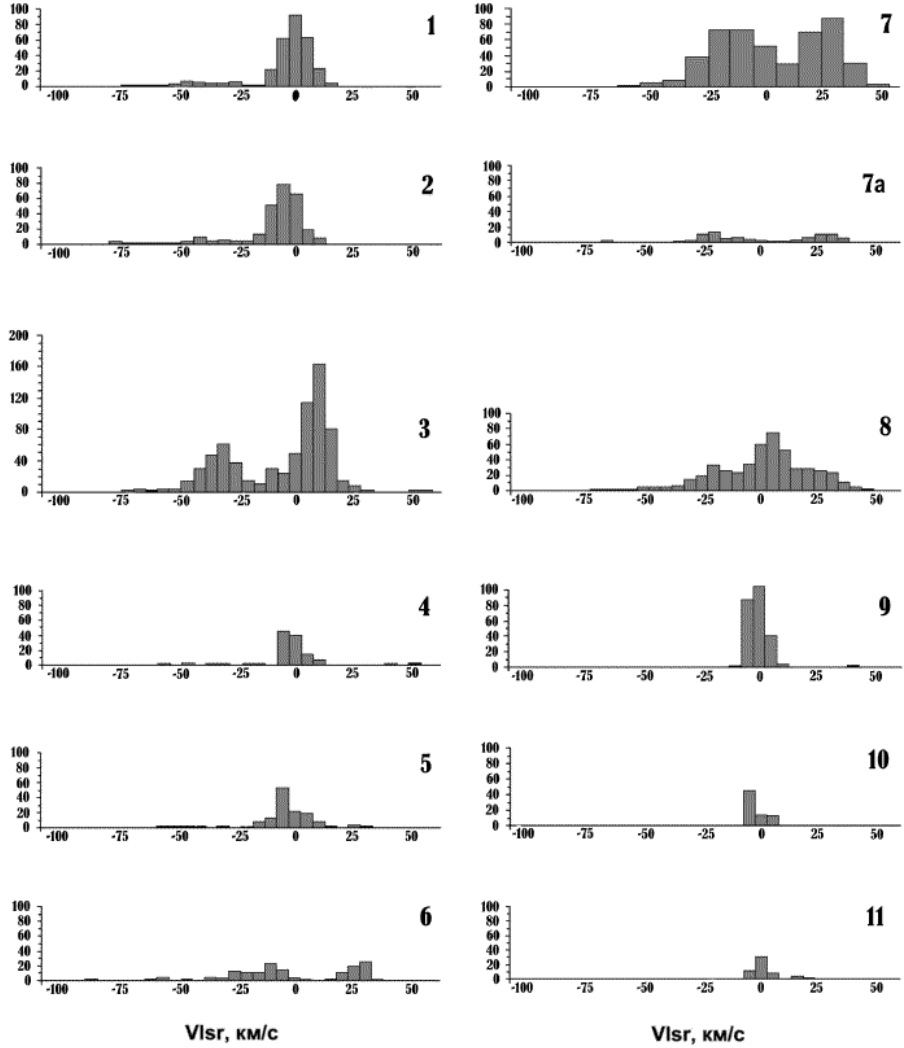}
%%% Figure:3
\caption{%
The velocity histograms as constructed from H$\alpha$ measurements for fields
nos.~1--11
shown in Fig.~2.}%
\end{figure*}

The velocity histograms shown in Fig.~3 were fitted with a set of Gaussians; the
fitting
results are presented in the table. The columns give, respectively, field
numbers and
boundary coordinates; mean radial velocities, FWHM and amplitude for each
Gaussian of
the histogram.

The results of our ionized-gas velocity measurements for the region can be
summarized as
follows.

The observed H$\alpha$ profiles in the Superbubble region in Cygnus consist
either of one
component, occasionally with a blue or red wing, or of two to three components.
In general, the intensity of the weak features observed in the wings does not
exceed 15--20$\%$ of the intensity of the main component.

Intense blueshifted and redshifted line components are observed toward several
areas
(fields nos.~3, 6, 7, 7a, and 8). In these fields, the shifted line components
are
systematically more intense than the main component.

According to the results in Fig.~3 and in the table, the peak velocities of the
intense
line components vary within the range $-40$ to $+45$~km~s$^{-1}$. Weaker
features in the
line wings at high positive and negative velocities are encountered in the range
$\approx -80$ to $\approx +55$~km~s$^{-1}$. (Features are also observed at
higher
positive and negative velocities but at a level below our signal-to-noise
ratio~$\ge 5$.)

\begin{table*}[t!]
\begin{center}
\caption{Radial velocity V(LSR) occurrence in the fields 1--11.}
\begin{tabular}{|c|c|c|c|c|c|}
\hline
Field& RA(2000):h,m,s     & D(2000):$^\circ$,'& V(max)      & HFWM       & A \\
\hline
1  & 20 18 15 - 20 22 15 & 40 10 - 41 10    & 0.15$\pm$ 0.03 & 13.8$\pm$ 0.1 &
91.4\\
   &                     &                  & -44.2$\pm$ 0.6 & 16.4$\pm$ 1.4 &
6.4 \\
   &                     &                  & -26.6$\pm$ 0.5 & 7.8$\pm$ 1.3  &
5.4 \\
\hline
2  & 20 22 15 - 20 28 15 & 40 07 - 40 30    &  -4.2$\pm$ 0.1 & 13.4$\pm$ 0.3 &
81.8 \\
   &                     &                  & -38.2$\pm$ 1.6 & 17.8$\pm$ 3.8 &
6.7 \\
\hline
3  & 20 25 15 - 20 30 30 & 40 30 - 41 35    &   9.3$\pm$ 0.2 & 11.5$\pm$ 0.3 &
161.7 \\
   &                     &                  & -31.7$\pm$ 0.3 & 16.6$\pm$ 0.8 &
58.2 \\
   &                     &                  &  -5.9$\pm$ 1.2 & 14.3$\pm$ 2.9 &
28.2 \\
\hline
4  & 20 15 30 - 20 18 30 & 41 25 - 42 10    &  -2.5$\pm$ 0.2 &  8.7$\pm$ 0.6 &
53.2 \\
\hline
5  & 20 25 15 - 20 27 45 & 41 35 - 42 00    &  -4.3$\pm$ 0.3 &  8.5$\pm$ 0.7 &
51.8 \\
\hline
6  & 20 24 45 - 20 26 30 & 42 00 - 42 25    & -13.1$\pm$ 1.1 & 23.3$\pm$ 2.5 &
17.1 \\
   &                     &                  &  27.4$\pm$ 0.5 & 10.2$\pm$ 1.2 &
25.8 \\
?  &                     &                  & -55.8$\pm$ 3.7 & 11.0$\pm$ 8.5 &
3.4 \\
\hline
7  & 20 24 30 - 20 29 50 & 42 30 - 43 20    &  27.2$\pm$ 1.0 & 20.2$\pm$ 0.6 &
93 \\
   &                     &                  & -13.5$\pm$ 0.7 & 31.7$\pm$ 1.0 &
79 \\
\hline
7a & 20 24 30 - 20 25 45 & 42 30 - 42 50    &  26.7$\pm$ 0.2 & 15.1$\pm$ 0.5 &
11 \\
   &                     &                  &  -9.2$\pm$ 0.9 & 15.0$\pm$ 2.0 &
5.5 \\
   &                     &                  & -22.4$\pm$ 0.2 &  8.5$\pm$ 0.6 &
14.2 \\
7a*& 20 24 30 - 20 25 45 & 42 30 - 42 50    &  26.4$\pm$ 0.4 & 17.3$\pm$ 1.0 &
16 \\
   &                     &                  & -19.7$\pm$ 0.3 & 15.2$\pm$ 0.5 &
21 \\
\hline
8  & 20 20 15 - 20 24 45 & 42 10 - 43 00    & -18.5$\pm$ 0.7 & 23.2$\pm$ 1.7 &
29 \\
   &                     &                  &   4.3$\pm$ 0.3 & 14.1$\pm$ 0.7 &
71 \\
   &                     &                  &  24.0$\pm$ 0.8 & 18.9$\pm$ 1.8 &
27 \\
\hline
9  & 20 54 00 - 20 04 00 & 42 30 - 46 00    &  -1.3$\pm$ 0.1 &  9.4$\pm$ 0.2 &
126 \\
\hline
10 & 20 48 00 - 20 55 00 & 43 15 - 44 35    &  -3.4$\pm$ 1.1 &   5$\pm$ 3    & 
60.4 \\
\hline
11 & 20 46 00 - 20 52 00 & 41 30 - 43 00    & -0.45$\pm$ 0.02 & 7.7$\pm$ 0.03& 
31.3 \\
   &                     &                  &  16.3$\pm$ 0.2  & 5.5$\pm$ 0.4 & 
3.5 \\
\hline

\end{tabular}
\end{center}
\end{table*}

It is of interest to compare our velocity measurements with the H$\alpha$
observations
by Reynolds (1983) with a Fabry--Perot spectrometer in the $l = 0$--240$^\circ$
band
along the Galactic plane at 2$^\circ$ intervals. These observations were carried
out with a low angular resolution (50$'$) but with a higher sensitivity than in
our measurements and a similar spectral resolution (12 km~s$^{-1}$). Reynolds
(1983) noted
the presence of weak emission features ($\approx3$~Rayleighs) at high positive
velocities forbidden by the differential Galactic rotation and associated them
with the
expansion of the X-ray Superbubble at a velocity from 20 to 50~km~s$^{-1}$. It
seems
that in Fig.~4 from the cited paper, two groups of forbidden positive
velocities,
$V(LSR)=+35$~km~s$^{-1}$ and $V(LSR)=+50\ldots52$~km~s$^{-1}$, can
be identified in the direction $l=70\ldots88^\circ$. The corresponding features
at negative
velocities are also seen in the range $l=72\ldots 82^\circ$, and these are also
represented by two groups, $V(LSR)=-20\ldots -22$~km~s$^{-1}$ and
$V(LSR)=-40\ldots-42$~km~s$^{-1}$.

\section{ANALYSIS OF THE OPTICAL, X-RAY, AND INFRARED OBSERVATIONS FOR THE
REGION:
THE ACTION OF Cyg~OB2 ON THE INTERSTELLAR MEDIUM}

In this section, using our interferometric observations and X-ray, infrared, and
radio
studies of the region, we attempt to answer the following questions:

Where is evidence for the action of the Cyg~OB2 wind on the interstellar gas?
Where is
the shell swept up by the wind? Where are observational manifestations of the
action of
intense ionizing radiation from Cyg~OB2 on the ambient gas? What are the
parameters of
the ambient interstellar medium that make the answers to these trivial questions
so
equivocal?

We immediately note that, first, the answers are not new but rest on a new line
of
reasoning, and, second, they require further observational confirmations.

The main difficulty involved in investigating the region is that the kinematic
distances
in this direction are poorly determined, because the line-of-sight projection of
the
Galactic rotation velocity changes little with distance. Therefore, the most
complex
question is whether the various gaseous and stellar components in this densely
populated
region along the Local spiral arm are physically related.

\subsection*{Kinematic Evidence of the Shell Swept up by the Wind}

The problem of revealing kinematic evidence for the existence of a single shell
swept up
by a strong wind from Cyg~OB2 is complicated by the fact that the observed
H$\alpha$
profile is the sum of emissions from the Local spiral arm up to distances of
3~kpc.
Consequently, the velocities can differ because of the different localization
of individual nebulae in distance and in the arm cross section. These
differences
attributable to the differential Galactic rotation and to the passage of spiral
density
waves lie within the range $-10\ldots+10$~km~s$^{-1}$ for the sky region in
question
(Sitnik \emph{et al.} 2001).

Even if we deal with a physical unity, then, since the hypothetical
shell is located in an inhomogeneous dense medium, the line velocity variations
over the image may result from the following:

(1) Small-scale gas density fluctuations, which cause a warping of the
ionization front
and, accordingly, variations in the observed radial velocity;

(2) Gas outflow via the champagne effect when the ionization front collides with
dense cloud cores; in the presence of several clouds, gas outflow is observed in
both
directions.

Since the kinematic effect of these factors does not significantly exceed the
speed
of sound in an ionized gas either, evidence for the action of the Cyg~OB2 wind
should be
sought by supersonic motions.

The weak features in the line wings observed at high positive and negative
velocities
are unreliable, because these are often identified at a noise level and may
result
from incomplete allowance for the radiation from faint stars. The high-velocity
motions
can also be associated with individual energy sources (the winds from WR and Of
stars and
blue supergiants or supernova explosions).

Thus, the intense shifted H$\alpha$ components that we detected in several
central
fields may serve as the only reliable kinematic evidence for the existence of a
shell.
It seems reasonable to attribute these intense high-velocity components to the
action of
a strong Cyg~OB2 wind. We emphasize that individual features at high negative
velocities can undoubtedly be produced by the radiation from distant nebulae in
the
Perseus arm. At the same time, the detected positive velocities reaching
$+55$~km~s$^{-1}$ have no obvious alternative explanation and most likely
suggest that
there are large-scale motions of ionized gas in the region.

If we deal with a physical unity, then the intense line features at high
positive and
negative velocities can characterize the receding and approaching sides of the
expanding
shell, respectively. Our interferometric observations (see Section~2) lead us to
conclude that the possible expansion velocity of the hypothetical ionized shell
reaches
$\approx 25$--50~km~s$^{-1}$.

A single-peaked H$\alpha$ line is emitted in the thin filaments that form the
giant ring
system. Its FWHM is 20--35~km~s$^{-1}$ for the group of filaments in the
northwest and
reaches 45~km~s$^{-1}$ in the southeast (the latter are considerably weaker).
The
results presented in Fig.~2 suggest that the velocity at the line peak in the
northwest
varies over the range $-1$ to $+6$~km~s$^{-1}$ (the rms error is
3--4~km~s$^{-1}$). In
the southeast, the velocity in the region of the weakest filaments varies over a
wider
range, from 4--6 to 20--22~km~s$^{-1}$, with an rms error of 5--9~km~s$^{-1}$.
Reynolds
(1983) measured the velocities of some filaments and found
 $V(LSR) = \pm 3$~km~s$^{-1}$, but these measurements were made with a clearly
insufficient angular
resolution of 50$'$ for such thin-filament structures.

Note that several authors (Chaffee and White 1982; Willson 1981; Piependbrink
and
Wendker 1988; Lockman 1989) observed the velocities of radio line features in
the range
$V(LSR) = -15 \ldots -20$~km~s$^{-1}$ to $V(LSR) = +15 \ldots
+20$~km~s$^{-1}$ toward the Cygnus~X complex by absorption and emission lines.
This is
generally considered as resulting from the different localization of the
emitting and
absorbing gas in distance rather than from peculiar motions. Gredel and Munch
(1994)
identified features in the interstellar C$_{2}$ absorption lines in the spectra
of the
Cyg~OB2 stars nos.~5 and~12 at velocities reaching $V(LSR) =
30$~km~s$^{-1}$,
which strongly suggests high-velocity motions of the molecular clumps in the
association
region. Interestingly, the two stars, which are 5~pc apart in the plane of the
sky,
exhibit absorption features at close velocities. In the opinion of the authors,
this suggests a layered structure of the absorbing molecular clumps.

\subsection*{Cyg~OB2: A New Estimate for the Mechanical Luminosity of the
Stellar Wind}

Previous studies of Cyg~OB2 have revealed an elliptical stellar structure,
$48'\times28'$
in size, with $\sim3000$~members. Three hundred OB stars, five O5 stars, the
only
northern-sky O3 star, three WR stars, and several O5~If, O4~IIIf, and LBV stars
were
identified In Cyg~OB2; the association includes the most luminous and most
massive stars
of the Galaxy (Reddish \emph{et al.} 1966; Torres-Dodgen \emph{et al.} 1991;
Massey and
Thompson 1991; Persi \emph{et al.} 1985). Reddish \emph{et al.} (1966) first
pointed out
that Cyg~OB2 differs from Galactic OB associations by compactness, large mass,
and high
star density, which make the object similar to young blue globular clusters in
the LMC.

Recently, a qualitatively new step has been made by Kn\"odlseder (2000). He
passed from
optical to infrared wavelengths, which allowed him to significantly increase the
stellar
population of the association by adding objects previously hidden due to strong
absorption: according to his estimates, $Av \simeq 5 - 20^{m}$ for Cyg~OB2
stars.
According to the new data, Cyg~OB2 is spherical in shape. The coordinates of its
center
are RA$(2000)=20^{\mathrm{h}} 33^{\mathrm{m}} 10^{s}$ and D$(2000)=41^\circ
12'$; $l =
80^\circ 7.2'$, $b= 0^\circ 43.8'$; the size of the region containing $90\%$ of
the
stellar population is 90$'$ (45~pc). In what follows, the distance to Cyg~OB2 is
assumed
to be $d=1.7$~kpc, as estimated by Massey and Thompson (1991).

The current mass of Cyg~OB2 is (4--10)$\times10^{4} M\odot$, the number of its
members
with spectral types earlier than~F3 reaches $8600\pm1300$, and the number of its
OB stars
is $2600\pm400$, including 120 O~stars. This mass and star density distinguish
Cyg~OB2
from other associations so clearly that Kn\"odlseder (2000) reached the
following
unequivocal conclusion: the object is actually the only Galactic representative
of the
population of young blue globular clusters, which are so numerous in the LMC.
The H--R
diagram for Cyg~OB2 clearly reveals main-sequence stars up to a mass of
85~M$\odot$;
although the association is compact, star formation in it was, probably, not
coeval
(Massey and Thompson 1991). The age of Cyg~OB2 is $3\times 10^{6}$ yrs, as
estimated by
Torres-Dodgen \emph{et al.} (1991); using new data on the stellar population,
Kn\"odlseder \emph{et al.} (2001) obtained $(2.5\pm0.1)\times 10^{6}$ yrs.

The wind mechanical luminosity for previously known members of the association
was
estimated to be L$_{w}\simeq 10^{38}$~erg~s$^{-1}$ (Abbott \emph{et al.} 1981).
The mass
loss rate for the five most luminous stars in Cyg~OB2 is $\sim$ (2--10)
$\times10^{-5}$
M$\odot$~yr$^{-1}$ at a wind velocity of 2000--7500~km~s$^{-1}$ (Persi \emph{et
al.}
1983; Leitherer \emph{et al.} 1982).

Given the new data on the Cyg~OB2 population, the above wind mechanical
luminosities
should be reestimated. We use a detailed analysis of the changes in wind energy
during
the evolution of a rich star grouping performed by Leitherer \emph{et al.}
(1992) (see
also Comeron and Torra 1994). As follows from the cited paper, at an early stage
($t\leq
2\times 10^{6}$ yrs) for normal metallicity, the wind mechanical luminosity is
kept
approximately constant, being $2\times10^{34}$~erg~s$^{-1}$ per unit total mass
of all
the stars with masses in the range 1--120~$M\odot$. The total mass for Cyg~OB2
is
$(4-10)\times10^{4}$~$M\odot$, which yields $Lw\simeq
(1-2)\times10^{39}$~erg~s$^{-1}$.
This mechanical luminosity is maintained at least for $2\times 10^{6}$ yrs and
increases
further as WR stars appear.

\subsection*{Possible Manifestations of the Shell Swept up by the Cyg~OB2 Wind}

The radius and expansion velocity of the shell that the Cyg~OB2 wind could
produce in a
medium with a known initial density over a cluster lifetime $ t\simeq
(2-3)\times 10^{6}$ yrs give the relations following from the classical
theory by
Castor \emph{et al.} (1975) and Weaver \emph{et al.} (1977):

%\begin{gather*}
$$
    R(t)= 66 n_{0}^{-1/5} L_{38}^{1/5} t_{6}^{3/5}\hspace{5pt} {\rm pc},
$$
%\end{gather*}
%\begin{gather*}
$$
   v(t)= 39 n_{0}^{-1/5}L_{38}^{1/5}t_{6}^{-2/5}\hspace{5pt}{\rm km}~{\rm
s}^{-1},
$$
%\end{gather*}
where $t_6$ is the age, Myr; $R$ is the radius, pc; and $L_{38}$ is the
stellar-wind
mechanical luminosity, $10^{38}$~erg~s$^{-1}$.

It follows from these relations that the possible expansion velocity of
$\approx50$~km~s$^{-1}$ determined from the optical emission can be produced by
the
Cyg~OB2 wind in a medium with a reasonable initial density n$_{0}\simeq
1$~cm$^{-3}$; in
this case, the shell radius can reach $\simeq 200$~pc.

The structures of such a size are represented by the giant system of optical
filaments
and by the X-ray Superbubble, which have been repeatedly discussed in the
literature.
Below, we briefly consider the main arguments for and against this
identification.

\textbf{The giant ring of optical filaments}. 

The giant ring of weak optical filaments
mentioned by Morgan \emph{et al.} (1955) and Struve (1957) has EM~$\leq
100$~cm$^{-6}$pc
(Dickel \emph{et al.} 1969). Ikhsanov (1960) considered this system of filaments
as a
physical unity that bounds the complex of nebulae ionized by Cyg~OB2 stars.

Based on the extinction $A_V$ estimated from the ratio of the H$\alpha$ and
radio
continuum intensities and on the distance dependence of $A_V$, Kapp-Herr and
Wendker
(1972) concluded that the filaments constituting the giant shell in the plane of
the sky
lay at different distances. Unfortunately, this method is unreliable in the
Cygnus
region, because the absorbing matter is inhomogeneous: we now know that $A_V$
varies
between 5$^{m}$ and 20$^{m}$ (Kn\"odlseder 2000) even toward the compact Cyg~OB2
grouping located at the same photometric distance of 1700~pc. Besides, in these
estimates, the optical emission was averaged over 11$'$ fields and the radio
emission
was estimated with a 30$'$ resolution, which is clearly not enough for such thin
filaments.

Our ionized-gas velocity measurements are consistent with the assumption that
the giant
system of filaments is the shell swept up by the Cyg~OB2 wind. We determined the
possible expansion velocity, $\approx50$~km~s$^{-1}$, from the H$\alpha$
emission of the
central region, mainly toward the Cygnus~X complex. The filaments with measured
velocities are mostly located on the Superbubble periphery and their radial
velocities
must be lower because of the projection effect.

The filaments in the southeast must be studied in more detail, because our
observations with poor spacial covering revealed here high positive
velocities, 20--22~km~s$^{-1}$. These are the weakest filaments among the
objects that we investigated toward Cygnus; the measurement error in the
velocity here is larger than that in the remaining region (5--9~km~s$^{-1}$;
see Section~3). Besides, it may well be that these filaments do not lie on
the shell periphery. As Fig.~1 shows, the system of optical filaments and
the X-ray Superbubble as a whole closely coincide in the plane of the sky,
but the X-ray region in the southeast goes outside the thin optical
filaments by 2--3$^\circ$. The high velocities of these filaments may also
be attributable to a local energy source.

Note that the ring structure of H~I clouds with the same angular size observed
in the
velocity range $-20$ to $+20$~km~s$^{-1}$ (Gosachinskii and Lozinskaya 1997),
probably,
corresponds, to the system of optical filaments. This is also an argument for a
physical unity.

\textbf{The X-ray Superbubble in Cygnus}. 

The X-ray Superbubble in Cygnus that was
identified by Cash \emph{et al.} (1980) using HEAO-1 observations with an
angular
resolution of $1^\circ .5\times3^\circ$ was considered by the authors
as a
physical unity associated with Cyg~OB2 and with the above system of optical
filaments.
The size of the Superbubble was estimated to be 450~pc, its X-ray luminosity is
$L_{\mathrm{X}}\simeq 5\times 10^{36}$~erg~s$^{-1}$, and the density of the
emitting
plasma is $n_{e}\approx 0.02$~cm$^{-3}$. Different authors explained the
formation of a
superbubble with such parameters either by sequential explosions of several tens
of
supernovae in Cyg~OB2 (Cash \emph{et al.} 1980), or by the combined action of
the wind
from the same association (Abbott \emph{et al.} 1981), or by a supernova
explosion
inside the previously formed low-density cavity (Higdon 1981), or by the
explosion of a
single supermassive star (Blinnikov \emph{et al.} 1982).

According to the alternative interpretation by Bochkarev and Sitnik (1985) and
Uyaniker
\emph{et al.} (2001), the X-ray Superbubble is not a physical unity but is the
sum of
emissions from supernova remnants, the shells around individual stars and
OB~associations swept up by the wind, cataclysmic variables, hot stellar
coronas, etc.
located in the Local spiral arm at distances from 0.5 to 4~kpc. The X-ray shell
morphology in the plane of the sky results from absorption of the X-ray emission
in a
dense layer of clouds at 700--900~pc (the so-called Great Cygnus Rift).

We performed a detailed analysis of the ROSAT observations in an effort to
reveal the
possible X-ray emission component produced by a strong Cyg~OB2 wind. Here, we
briefly
summarize the main conclusions of our analysis.

The fact that the shell morphology of the Superbubble is mainly explained by
absorption
may be considered to have been firmly established.

The ROSAT observations confirm the previous conclusion by Bochkarev and Sitnik
(1985)
that most of the X-ray emission (up to 50--80$\%$) is produced by point sources
of a
different nature. However, their conclusion that the diffuse part of the X-ray
emission
originates from the cavities swept up by individual OB~associations in the
Cygnus region
is apparently not confirmed. A simple superposition of the association
boundaries on the
ROSAT X-ray map reveals no one-to-one coincidence and this cannot be explained
by
absorption [see Fig.~1 in Uyaniker \emph{et al.} (2001)].

\begin{figure*}[t!]
\centering\includegraphics[width=14.cm]{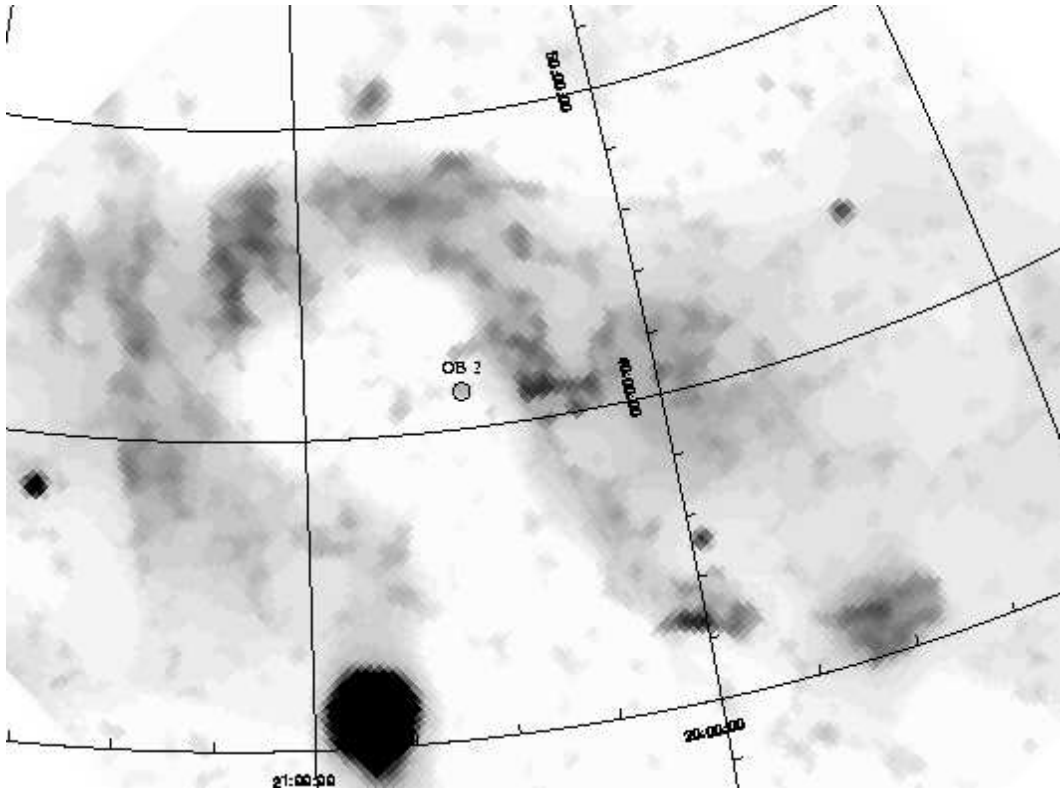}
%%% Figure:4
\caption{%
The intensity distribution of the diffuse X-ray emission, as constructed from
ROSAT data in the energy range 0.7--0.9 keV . The point sources were removed; the
bright
spot in the south is the supernova remnant  Cygnus Loop.}%
%\label{%
%}%
\end{figure*}

The conclusion by Uyaniker \emph{et al.} (2001) that individual regions of the
Cygnus
Superbubble have different radiation spectra and cannot be considered as a
physical
unity is based on a comparison of the X-ray hardness and the column density of
the
absorbing matter $N$(H~I). The authors pointed out that the bulk of the
absorbing matter
in the fields with the strongest absorption have high negative velocities,
$V(LSR)\leq-30\ldots -40$~km~s$^{-1}$. This result is of great interest
but it
requires a further detailed analysis. Indeed, the above effect implies that
either the
bulk of the matter lies far and cannot absorb the X-ray emission or it is
located in the
region under consideration and has been affected by a shock wave, which accounts
for the
observed high negative velocity. (If this is the case, then the velocity of the
absorbing neutral hydrogen closely matches our derived expansion velocity of the
hypothetical shell.)

In any case, however, the X-ray spectra of different parts of the Superbubble
should be
compared only after removing the point sources whose contribution is comparable
to the
diffuse component.

\textbf{The diffuse X-ray emission from the Superbubble}. To identify the
diffuse X-ray
emission component, we used the ROSAT XRT/PSPC all-sky survey maps cleaned from
point
sources; the angular resolution is 12$'$ (Snowden \emph{et al.} 1997).

The derived intensity distribution of the diffuse emission in the range 
0.7--0.9~keV  is shown
in Fig.~4. The total size of the X-ray region determined from the outermost
areas of the
Superbubble reaches 15--18$^\circ$.

An extended arc-shaped X-ray region, which forms the eastern and southeastern
parts of
the Superbubble ($RA= 21^{\mathrm{h}} 0^{\mathrm{m}} \ldots 21^{\mathrm{h}}
33^{\mathrm{m}}$; $D = 31^\circ \ldots 46^\circ$), is clearly seen in the
figure. This
eastern X-ray arc currently seems to be the most plausible observational
manifestation of
the part of the shell swept up by the Cyg~OB2 wind in a tenuous interstellar
medium.

According to Uyaniker \emph{et al.} (2001), this arc structure (the southern arc
in the Galactic coordinate system used in their paper) was produced by the
Cyg~OB4 wind. This
conclusion does not seem justifiable. Indeed, Cyg~OB4 is the poorest association
in the region; it consists of a mere two OB stars (B9~Iab and B0.5~IV; see Bl ha
and
Humphreys
1989). Besides, the elongated shape of the X-ray arc reveals no clear
morphological
relationship to this association.

Figure~5a shows the intensity distribution of the diffuse X-ray emission, as
constructed
from ROSAT data, superimposed on the distribution of the intensity ratio of
infrared
emissions in two bands (60/100~mkm), as constructed from IRAS data. Only the
brightest
X-ray isophotes (confidence level 100$\sigma$) corresponding to 16, 32, 100,
200, and
13000 in units of $10^{-6}$ counts~s$^{-1}$ per square arcminute or to 4, 8, 26,
51, 3300 in
units of $10^{-12}$ erg~s$^{-1}$ per square arcminute are shown.

The dust temperature distribution in the intensity ratio of infrared 60/100~mkm
emissions (see Fig.~5) reveals an extended ($\approx15^\circ$) elliptical
structure of
warmer dust that generally copies the shape of the X-ray Superbubble and
surrounds it.
An extended arc is also clearly seen here east of the region. This arc of warm
dust has
the same shape as the X-ray arc, and it also surrounds the latter from outside.
Such a
region of warm dust can be produced by inelastic collisions of dust grains with
hot X-ray plasma, as is the case in supernova remnants.

Figure~5b shows the temperature distribution of the X-ray emitting plasma that
we
derived from ROSAT data. The temperature map was constructed from calibrations
of the
temperature and absorption effects on the fluxes in three spectral bands of the
PSPC/ROSAT detector (Finoguenov 1997) and superimposed on the intensity
distribution of
the diffuse X-ray emission. The temperature map was obtained from the so-called
spectral
hardness, which is defined as the ratio $(F2-F1)/(F2+F1)$, where $F1$ is the
0.5--0.9~keV flux and $F2$ is the 0.9--2~keV flux, and smoothed with a Gaussian
with the
FWHM $\sigma=4$~pixels. The white spot inside the Superbubble may result from
the
absorption corresponding to a column density of absorbing atoms $
N(\mathrm{HI})\geq
4\times 10^{21}$~cm$^{-2}$.

A detailed analysis of the distributions of temperature, intensity of the
diffuse X-ray
emission, and absorbing atoms on the line of sight is beyond the scope of this
study.
Here, it it only important to note that the X-ray temperature differences over
the
entire length of the eastern arc are small, mainly from 0.50 to 0.57~keV (except
for the
compact source). This allows the eastern arc to be considered as a physical
entity,
which most likely represents the X-ray emission from the part of the shell swept
up by the Cyg~OB2 wind.

The X-ray emission from the plasma heated by the Cyg~OB2 wind, probably,
contributes
significantly to other parts of the Cygnus Superbubble as well. However, it
cannot yet be
reliably identified against the background of the total emission along the Local
spiral
arm. Therefore, we estimate the parameters of the entire X-ray emission as an
upper
limit on the luminosity and mass of the hot plasma attributable to the Cyg~OB2
wind.

The identified diffuse X-ray 0.4--2.1-keV emission from the entire Superbubble
corresponds to 275.1~counts~s$^{-1}$. For an equilibrium plasma of solar
chemical
composition at the temperature $kT_{e} = 0.6$~keV and a distance of 1.7~kpc,
this gives a
bolometric luminosity L$_{\mathrm{X}}\simeq 1.2\times 10^{36}$~erg~s$^{-1}$. In
this
case, most of the luminosity is emitted in the 0.4--2.1~keV band
(L$_{0.4-2.1}\simeq 0.88\times 10^{36}$~erg~s$^{-1}$).

The mass of the X-ray emitting plasma reaches M$\simeq 8.1\times
10^{4}$~M$_\odot$, and
the electron density is $n_{\mathrm{e}}\simeq 4.6\times 10^{-3}$~cm$^{-3}$.

We emphasize once again that this is an upper limit on the luminosity and mass
of the
hot plasma, which may be attributable to the Cyg~OB2 wind. However, even if most
of the
emission from the X-ray Superbubble is attributable to the Cyg~OB2 wind, our new
estimate
of the wind mechanical luminosity suggests that there are currently no problems
with the
Superbubble energetics.

\subsection*{The Interstellar Medium Around Cyg~OB2: A General Scheme}

The last question posed at the beginning of this section seems easiest to
answer: the
shell swept up by the wind can be hidden in a highly inhomogeneous interstellar
medium
characterized by dense clouds with even denser compact clumps and a tenuous
intercloud
medium.

\begin{figure*}[t!]
\centering\includegraphics[width=11.cm]{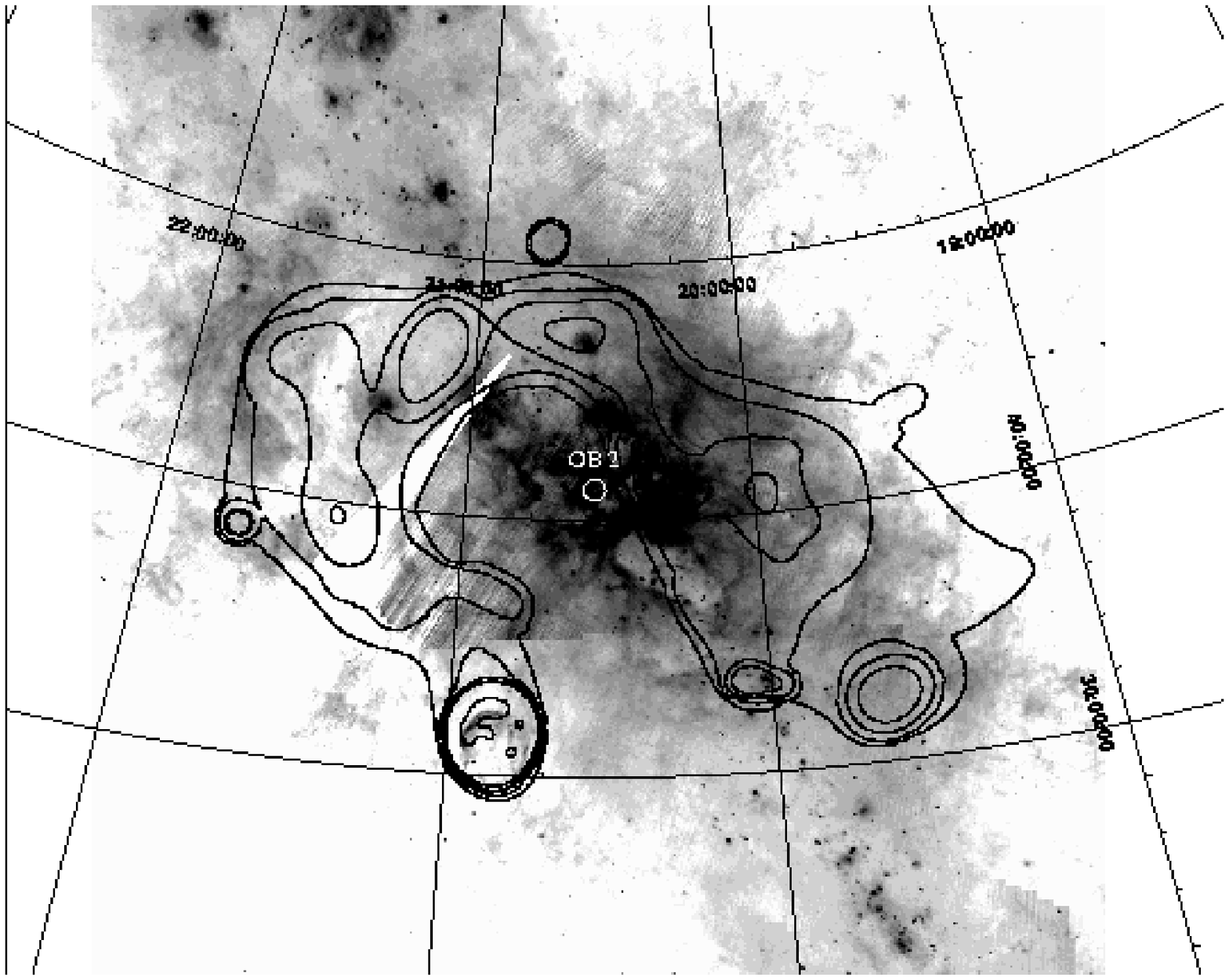}

\centering\includegraphics[width=11.cm]{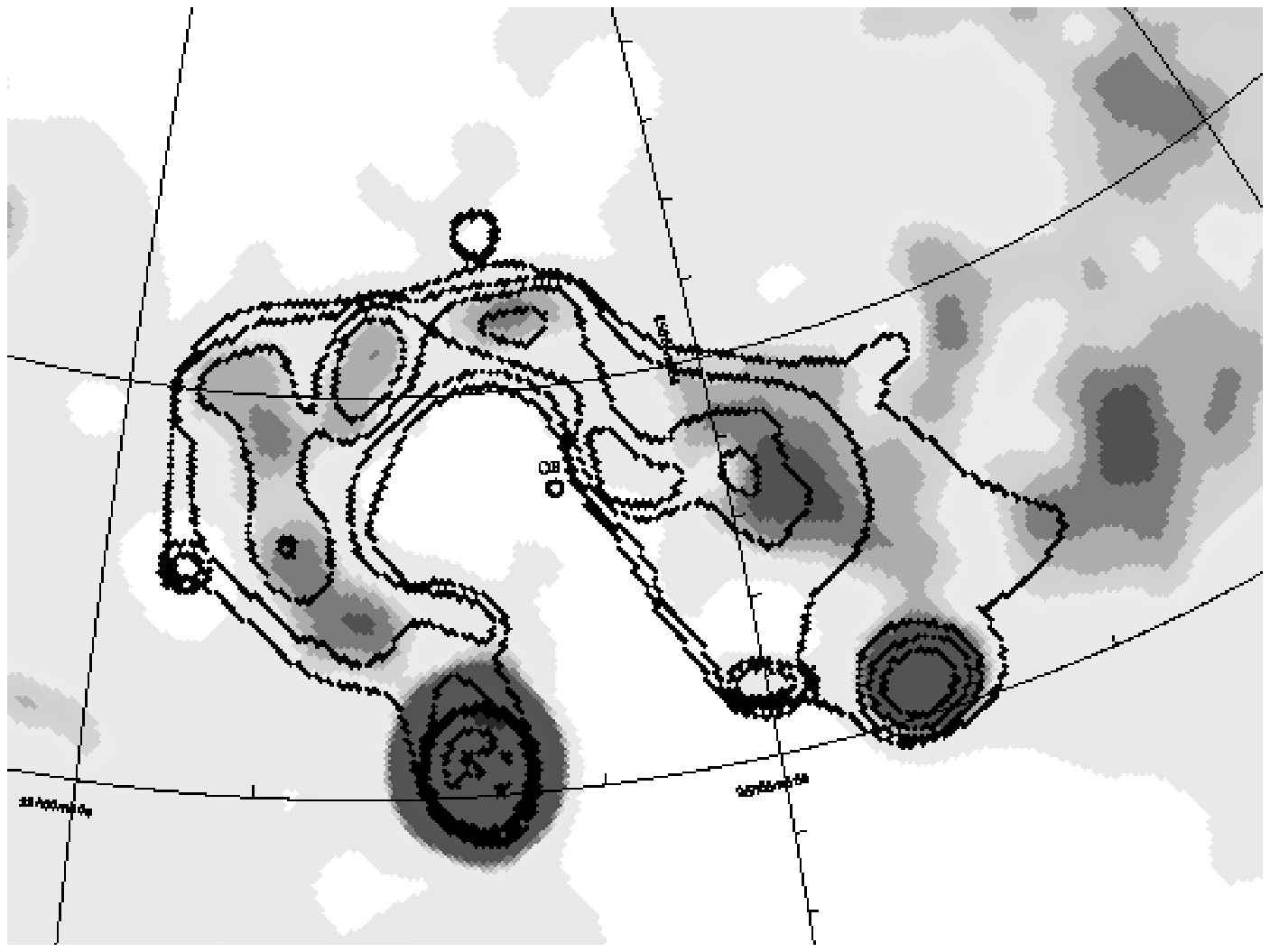}
%%% Figure:5
\caption{%
(a) The intensity distribution of the diffuse X-ray 0.7--0.9~keV emission
(indicated by
isophotes) superimposed on the distribution of the intensity ratio of infrared
emissions
in two bands ($60/100$~mkm), as constructed from IRAS data (indicated by shades
of gray). The brightest X-ray isophotes (confidence level 100$\sigma$)
corresponding to 4,
8, 26, 51, and 3300 in units of $10^{-12}$~erg~s$^{-1}$~arcmin$^{-2}$ are
presented. The
position of Cyg~OB2 is marked. (b) The X-ray plasma temperature distribution
(indicated
by shades of gray) superimposed on the same X-ray intensity distribution. The
temperature levels, from darkest (colder plasma) to lightest (hotter plasma),
correspond
to 0.24, 0.34,  0.45, 0.50, 0.54, 0.57, 0.63, and 0.85 keV.}%
%\label{%
%}%
\end{figure*}

The conclusion of a high local gas density in the Cyg~OB2 region follows from an
analysis of the interstellar extinction toward this region.

According to Neckel and Klare (1980), the extinction toward Cyg~OB2 reaches
A$_{V}=7^{\mathrm{m}}$ at a distance of $\approx1$~kpc; in nearby fields,
A$_{V}=7^{\mathrm{m}}$ is reached at a distance of 1--2~kpc. At larger
distances, the
extinction increases only slightly up to a limiting distance of 3~kpc. According
to
Ikhsanov (1959), in the extended region bordering Cyg~OB2 in the west, north,
and south,
the extinction monotonically increases to $A_V\simeq 2^{m}.6$ up to a
distance
of 1.15--1.4~kpc and then sharply increases to $A_V\simeq 6.5-7^{m}$ at a
distance of 2.2--3.5~kpc. The brightness ratio of optical nebulae in the radio
continuum
and H$\alpha$ toward Cyg~OB2 gives an extinction $A_V = 4-6^{m}$ (Dickel
and
Wendker 1978). Thus, the extinction in the entire extended region up to a
distance of 3~kpc does not significantly exceed $A_V\simeq 7^{m}$.

At the same time, the extinction derived from photometry of Cyg~OB2 stars is
A$_{V}=
5\ldots 20^{\mathrm{m}}$ (Kn\"odlseder 2000). Hence, a significant fraction of
the
absorbing matter (up to $50\%$ of the column density) is local relative to
Cyg~OB2.

The conclusion of a high local gas density in the Cyg~OB2 region can also be
drawn from
other considerations. The star density at the center of the young globular
cluster
Cyg~OB2 is 40--150~M$\odot$/pc$^{-3}$, as estimated by Kn\"odlseder (2000).
Since only
part of the mass of the parent molecular cloud turns into stars, we obtain a
lower limit
on the initial atomic hydrogen density in the parent cloud: n$(H_{2}) \geq
(1-3)\times 10^{3}$~cm$^{-3}$.

The dense matter of the parent molecular cloud left after the main
star-formation
episode is rapidly removed from the central region of the globular cluster by
ionizing
radiation and winds from the formed stars. However, the densest gaseous clumps
are not
swept up by the shock wave but are compressed and remain immersed in a tenuous
cavity.
Such dense molecular clumps live long enough and can presently be observed in
the
Cyg~OB2 region. They are actually observed: the clumpy structure of the
molecular gas
composed of separate clumps was revealed by interstellar C$_{2}$ absorption
lines in the
spectra of the Cyg~OB2 stars nos.~5 and~12 (Gredel and Munch 1994). The
characteristic
size of the clumps is $\leq 5$~pc (the most probable size is 1~pc) and their
mass is
$\approx200$~ M$\odot$; the clumps are located in the Cyg~OB2 region. Using the
size and
column density $N(\mathrm{C}_{2})$, we obtain the molecular hydrogen density in
the
clumps, $n(\mathrm{H}_{2}) \sim 100 \ldots 10^{3}$~cm$^{-3}$. (The lower limit
at
$l=5$~pc is $n(\mathrm{H}_{2}) = 300\ldots 600$~cm$^{-3}$ for star no.~12 and
$n(\mathrm{H}_{2})= 130\ldots 300$~cm$^{-3}$ for star no.~5). Such dense clumps
are
immersed in a low-density gas.

\textbf{Cygnus~• as an H~II region ionized by Cyg~OB2}. 

The Lyman continuum flux from
120 O-stars of Cyg~OB2 alone is $\simeq 10^{51}$ phot.~s$^{-1}$ (Kn\"odlseder
2000). In
a homogeneous medium with a density of 1, 10, and 100~cm$^{-3}$, the radius of
the
corresponding H~II region reaches 300, 65, and 15~pc and the emission measure is
300,
6500, and 150~000~cm$^{-6}$pc, respectively.

Whereas the absence of intense optical emission from this ionized gas can be
explained by
strong absorption, its intense thermal radio emission must be observed in any
case.
Using the exclusion method, we find only one possibility: this emission is
represented
by the Cygnus~X complex of radio sources. [This possibility has been repeatedly
discussed
starting from Veron (1965); see also Landecker 1984]. In any case, the regions
of ionized
gas with appropriate density and emission measure are observed only in Cygnus~X.
The
bright features of the complex are observed against the background of an
extended weaker
radio continuum source, 6--7$^\circ$ in size, with Cyg~OB2 being projected onto
its
central part (Wendker 1970; Huchmeier and Wendker 1977). This diffuse source was
immediately identified by the authors as an H~II region ionized by Cyg~OB2 (see
also
Wendker \emph{et al.} 1991). If the extended source is at the distance of
Cyg~OB2, then
its radius reaches 100~pc, EM~$\simeq 2000$cm${-6}$~pc, and the mean electron
density is
$\approx3$~cm$^{-3}$. EM~$\simeq 1000\ldots 50000$~cm$^{-6}$pc in the bright
H~II regions
of Cygnus~X and reaches $10^{6}$--$10^{7}$~cm$^{-6}$pc in compact radio clumps;
the
density is $n_{e} = 10-100$~á¬$^{-3}$ in the bright radio sources and at
least
an order of magnitude higher in compact clumps (see, e.g., Dickel \emph{et al.}
1969;
Piependbrink and Wendker 1988).

Thus, the parameters of the Cygnus~X complex agree with those expected for the
H~II
region ionized by Cyg~OB2 stars. The only objection to this identification is
the
long-standing and popular belief that Cygnus~X is not a physical unity but is
the sum of
emissions from the many sources located at different distances, from 500 pc to
3--4~kpc,
along the spiral arm (see, e.g., Wendker 1984; Piependbrink and Wendker 1988;
and
references therein).

The following two basic methods of distance determination underlie this belief:
from the
radial velocities of sources and from the extinction $A_{\mathrm{V}}$. Both are
unreliable in the direction of Cygnus (see above).

Therefore, we cannot rule out the possibility that the diffuse component and
some of the
sources forming the Cygnus~X complex are located at the same distance of
1700~pc. Of
course, individual features of the complex can lie nearer or farther. In this
case, the
Cygnus~X complex can be a remnant of the parent cloud ionized by Cyg~OB2 stars.
[This
conclusion was also drawn by Landecker (1984) from detailed measurements of the
complex
in the H166$\alpha$ recombination line.] Separate dense clumps of this cloud
became
nests for the ongoing star formation and have their own, local ionization
sources.

A number of findings argue for the validity of this assumption.

\textbf{The dust temperature distribution in the region}. Figure~5 shows the
distribution of the intensity ratio of infrared emissions in two bands
(60/100~mkm), as
constructed from IRAS data, which reflects the dust temperature distribution in
the
entire region under study. The highest temperature in a large region containing
the
Cygnus Superbubble is observed toward the Cygnus~X complex. The Cyg~OB2 cluster
lies at
the center of the region of warm dust, and the warmest local spots are directly
projected
onto Cyg~OB2 and surround it. This is evidence of the physical interaction
between
Cyg~OB2 stars and a dense cloud.

\textbf{The CO distribution in the region}. The Cyg~OB2 cluster lies within a
local
region of reduced brightness surrounded by the brightest CO~clouds in the
direction of
Cygnus~X. This is clearly shown by the CO isophotal maps integrated over the
velocity
range $-25$ to $+25$~km~s$^{-1}$ and over narrower velocity ranges: $-5$--0;
0--5,
5--10, and 10--15~km~s$^{-1}$ [see Leung and Thaddeus (1992) and references
therein].
This is consistent with the proposed model, because the parent molecular cloud
must be
destroyed in close proximity to Cyg~OB2.

\textbf{The H~I distribution in the Cygnus~X region}. Note also the shell
structure of
H~I and CO clouds identified by Gosachinskii \emph{et al.} (1999) around the
Cygnus~X
complex. Although this extended shell was identified with major reservations,
its
existence, if confirmed, argues for a single complex. These authors (see also
Gosachinskii and Lozinskaya 1997) also pointed out that the coordinates and
radial
velocities of H~II regions, as inferred from the observations by Piependbrink
and
Wendker(1988) in radio recombination lines, create a semblance of a ring
structure,
suggesting a large-scale radial motion in Cygnus~X in the velocity range $-12$
to
$+10$~km~s$^{-1}$.

\textbf{A general scheme for the interaction of Cyg~OB2 with the interstellar
medium}

Currently available optical, X-ray, infrared, and radio observations allow
the following general scheme of the region to be considered.

The blue globular cluster Cyg~OB2 was formed in a medium with a high initial
density and mainly destroyed the parent molecular cloud over its
lifetime. The radio emission from the remnants of the parent cloud ionized
by Cyg~OB2 is represented by the diffuse component and, possibly, by
individual sources of the Cygnus~X complex. The densest compact clumps of
the destroyed molecular cloud are observed in radio lines in absorption.

The high gas velocities that we detected in optical nebulae are attributable
to an expansion of the shell swept up by the Cyg~OB2 wind. The expansion
velocity of the swept-up shell reaches 25--50~km~s$^{-1}$; this shell is
bounded by a giant system of thin filaments.

The burst of the Cyg~OB2 wind into a low-density intercloud medium is
responsible for the formation of an extended region of diffuse X-ray
emission. The cluster was apparently formed not at the center but at the
edge of a dense cloud. Therefore, the burst of the wind into a low-density
medium in the southeastern direction produced the most prominent arc-shaped
structure of X-ray and infrared emission. This arc-shaped structure seems to
be the most plausible manifestation of the shell swept up by a strong
Cyg~OB2 wind. That is why Cyg~OB2 and the dense ionized cloud (Cygnus~X) are
located not at the center but on the northwestern boundary of the extended
X-ray region (the Cygnus Superbubble).

\section*{CONCLUSIONS}

Our most conservative conclusions can be formulated as follows:

(1) Our interferometric H$\alpha$ observations have revealed high-velocity
motions of ionized gas, which may be related to an expansion of the
hypothetical shell swept up by the Cyg~OB2 wind in a medium with an initial
density of 1--10 cm$^{-3}$. The expansion velocity determined by these
motions reaches 25--50~km~s$^{-1}$.

(2) Given the number of OB~stars increased by an order of magnitude, Cyg~OB2
has the strongest stellar wind among Galactic associations: according to our
new estimate, $L_{\mathrm{W}} \simeq (1-2)\times10^{39}$~erg~s$^{-1}$. The
wind mechanical luminosity over a Cyg~OB2 lifetime of $ \approx (2-3)\times
10^{6}$ yrs is high enough to produce a shell comparable in size to the
X-ray Superbubble and to the giant system of optical filaments.

(3) The most plausible manifestations of the shell swept up by the Cyg~OB2
wind in a tenuous medium are the following: the giant system of optical
filaments; the diffuse component of the X-ray Superbubble in Cygnus,
primarily the extended eastern arc that forms it; the extended elliptical
region of warm dust, in particular, the eastern arc of warm dust that bounds
the arc-shaped X-ray region.

(4) The Cyg~OB2 cluster formed in a medium with a high initial density. The
remnants of the parent molecular cloud ionized by Cyg~OB2 stars are,
probably, represented by the diffuse component and, possibly, by individual
sources of the Cygnus~X complex.

The proposed scheme undoubtedly requires a thorough observational
verification.  Of greatest interest are searches for high-velocity features
of interstellar absorption lines in the entire region; a detailed analysis
of the spectrum for the diffuse X-ray emission component of the Superbubble;
and a comparison of the column density of the absorbing gas derived from the
X-ray spectrum with direct CO line measurements.  Such a study is already
under way.

\section*{ACKNOWLEDGMENTS}

%\bigskip
%\bigskip

This work was supported by the Russian Foundation for Basic Research (project
no.~01-02-16118) and the Program ``Astronomy''  (project no.~1.3.1.2).

\section*{REFERENCES}

\begin{enumerate}
    \parindent=0pt \parskip=0pt \parsep=1pt
    \bibindent=0pt
    \frenchspacing    % AV - to get right spacing after ``et al.''
    \hyphenpenalty=10000
    \itemindent=-10pt   
    \itemsep=0pt                            %
    \listparindent=0pt                      %
    \labelsep=0pt                           %
    \leftmargin=1.0em
    \advance\leftmargin\labelsep
  \sloppy\clubpenalty4000\widowpenalty4000%

\item
D.~C.~Abbott, J.~H.~Bieging, and E.~Churchwell, Astrophys. J. {\bf 250}, 645
(1981).

\item
C.~Blaha and R.~Humphreys, Astron. J. {\bf 98}, 1598 (1989).

\item
S. I. Blinnikov, V. S. Imshennik, and V. P. Utrobin, Pis'ma Astron. Zh.
{\bf 8}, 671 (1982)  [Sov. Astron. Lett. {\bf 8}, 361 (1982)].

\item
N. G. Bochkarev and T. G. Sitnik, Astrophys. Space Sci. {\bf 108}, 237
(1985).

\item
P.~Brand and W.~J.~Zealey, Astron. Astrophys. {\bf 38}, 363 (1975).

\item
W.~Cash, P.~Charles, S.~Bowyer, \emph{et al.}, Astrophys. J. Lett. {\bf 238},
71 (1980).

\item
J.~Castor, R.~McCray, and R. Weaver, Astrophys. J. Lett. {\bf 200}, 107
(1975).

\item
F.~H~Chaffee and R.~E.~White, Astrophys. J., Suppl. Ser. {\bf 50}, 169
(1982).

\item
F.~Comeron and J.~Torra, Astrophys. J. {\bf 423}, 652 (1994).

\item
H.~R.~Dickel and H.~J.~Wendker, Astron. Astrophys. {\bf 66}, 289 (1978)

\item
H.~H.~R.~Dickel, H.~Wendker, and J.~H.~Bieritz, Astron. Astrophys. {\bf 1},
270 (1969).

\item
A. V. Finoguenov, Candidate's Dissertation (Inst. Kosm. Issled. Ross. Akad.
Nauk, Moscow, 1997).

\item
I. V. Gosachinskii and T. A. Lozinskaya, Astron. Zh. {\bf 74}, 201 (1997)
[Astron. Lett. {\bf 41}, 174 (1997)].

\item
I. V. Gosachinskii, T. A. Lozinskaya, and V. V. Pravdikova, Astron. Zh.
{\bf 76}, 453 (1999)  [Astron. Rep. {\bf 43}, 391 (1999)].

\item
R.~Gredel and G.~Munch, Astron. Astrophys. {\bf 285}, 640 (1994).

\item
J.~C.~Higdon, Astrophys. J. {\bf 244}, 88 (1981).

\item
W.~K.~Huchmeier and H.~J.~Wendker, Astron. Astrophys. {\bf 58}, 197 (1977).

\item
R. N. Ikhsanov, Izv. Krym. Astrofiz. Obs. {\bf 21}, 257 (1959).

\item
R. R. Ikhsanov, Astron. Zh. {\bf 37}, 988 (1960) [Sov.
Astron. {\bf 4}, 923 (1960)].

\item
A.~V.~Kapp-Herr and H.~J.~Wendker, Astron. Astrophys. {\bf 20}, 313 (1972).

\item
J.~Knodlseder, Astron. Astrophys. {\bf 360}, 539 (2000).

\item
J.~Knodlseder, M.~Cervino, D.~Schaerer, \emph{et al.}, astro-ph/0104074 (2001).

\item
T.~L.~Landecker, Astron. J. {\bf 89}, 95 (1984).

\item
C.~Leitherer, H.~Hefele, O.~Stahl, and B.~Wolf, Astron. Astrophys. {\bf 108},
102 (1982).

\item
C.~Leitherer, C.~Roberts, and L.~Drissen, Astrophys. J. {\bf 401}, 594
(1992).

\item
E.~O.~Leung and P.~Thaddeus, Astrophys. J., Suppl. Ser. {\bf 81}, 267 (1992).

\item
F.~Lockman, Astrophys. J., Suppl. Ser. {\bf 70}, 469 (1989).

\item
P.~Massey and A.~B.~Thompson, Astron. J. {\bf 101}, 1408 (1991).

\item
W.~W.~Morgan, B.~Stromgren, and H.~M.~Johnson, Astrophys. J. {\bf 121}, 611
(1955).

\item
Th.~Neckel and G.~Klare, Astron. Astrophys., Suppl. Ser. {\bf 42}, 251
(1980).

\item
P.~Persi, M.~Ferrari-Toniolo, and G.~L.~Grasdalen, Astrophys. J. {\bf 269},
625 (1983).

\item
P.~Persi, M.~Ferrari-Toniolo, M.~Tapia, \emph{et al.}, Astron. Astrophys.
{\bf 142}, 263 (1985).

\item
A.~Piependbrink and H.~J.~Wendker, Astron. Astrophys. {\bf 191}, 313 (1988).

\item
V. V. Pravdikova, Pis'ma Astron. Zh. {\bf 21}, 453 (1995)
[Astron. Lett. {\bf 21}, 403 (1995)].

\item
V.~L.~Reddish, L.~Lawrens, and N.~M.~Pratt, Publ. R. Obs. Edinb. {\bf 5}, 111
(1966).

\item
R.~J.~Reynolds, Astrophys. J. {\bf 268}, 698 (1983).

\item
E. G. Sitnik, A. M. Mel'nik, and V. V. Pravdikova, Astron. Zh. {\bf 78}, 40
(2001)  [Astron. Rep. {\bf 45}, 34 (2001)].

\item
S.~L.~Snowden, E.~Egger, M.~J.~Freyberg, \emph{et al.}, Astrophys. J.
{\bf 485}, 125 (1997).

\item
O.~Struve, Sky Telesc. {\bf 16}, 118 (1957).

\item
A.~V.~Torres-Dodgen, M.~Tapia, and M.~Carroll, Mon. Not. R. Astron. Soc.
{\bf 249}, 1 (1991).

\item
B.~Uyaniker, E.~Furst, W.~Reich, \emph{et al.}, Astron. Astrophys. {\bf 371},
675 (2001).

\item
P.~Veron, Ann. Astrophys. {\bf 28}, 391 (1965).

\item
R.~Weaver, R.~McCray, J.~Castor, \emph{et al.}, Astrophys. J. {\bf 218}, 377
(1977).

\item
H.~J.~Wendker, Astron. Astrophys. {\bf 4}, 378 (1970).

\item
H.~J.~Wendker, Astron. Astrophys., Suppl. Ser. {\bf 58}, 291 (1984).

\item
H.~J.~Wendker, L.~A.~Higgs, and T.~L.~Landecker, Astron. Astrophys.
{\bf 241}, 551 (1991).

\item
R.~F.~Willson, Astrophys. J. {\bf 247}, 116 (1981).

\end{enumerate}

\hfill{\it Translated by V. Astakhov}

\end{document}